\documentstyle[aps,multicol,epsfig]{revtex}
\begin{document}
\hbadness=10000
\title{Motion of vortex lines in nonlinear wave mechanics}
\author{Iwo Bialynicki-Birula$^{1,4}$\footnote{email: birula@cft.edu.pl},
Zofia Bialynicka-Birula$^{2,3}$}
\address{Center for Theoretical Physics$^1$, Institute of Physics,
Polish Academy of Sciences$^2$, College of Science$^3$,\\ Al. Lotnik\'ow 32/46, 02-668
Warsaw, Poland,\\ Institute of Theoretical Physics$^4$, Warsaw University, Ho\.za 69,
00-681 Warsaw, Poland}

\maketitle
\begin{abstract}
We extend our previous analysis of the motion of vortex lines [I. Bialynicki-Birula, Z.
Bialynicka-Birula and C. \'Sliwa, Phys. Rev. A {\bf 61}, 032110 (2000)] from linear to a
nonlinear Schr\"odinger equation with harmonic forces. We also argue that under certain
conditions the influence of the contact nonlinearity on the motion of vortex lines is
negligible. The present analysis adds new weight to our previous conjecture that the
topological features of vortex dynamics are to a large extent universal.
\end{abstract}

\begin{multicols}{2}

\section{Introduction}

Owing to recent advances in the experimental studies of Bose-Einstein condensation (BEC),
a detailed theoretical description of the motion of vortex lines in the condensate may
soon become testable. In this report we continue our previous analysis \cite{bbs} of the
evolution of vortex lines as determined by the quantum mechanical wave equations. We were
able to extend our methods of obtaining analytic solutions to a nonlinear Schr\"odinger
equation.

The standard tool in the study of BEC is the Gross-Pitaevski\v{\i} (GP) equation that
describes zero range interactions of the condensate particles. All studies of the
condensate dynamics based on this equation must rely either on numerical methods or on
approximations \cite{sf}, since analytic solutions of the GP equation in three dimensions
are not available. In order to obtain a rich family of {\em explicit} solutions allowing
for elaborate vortex line structure, we modify the form of nonlinearity replacing the
zero range interaction by harmonic forces. We are fully aware that atoms do not interact
via harmonic forces but the aim of our study was to answer the question: Does a nonlinear
term modify in an essential way our results obtained for the linear Schr\"odinger
equation? In Section II we prove that the answer to this question is negative in the case
of nonlinearity corresponding to harmonic forces. In the other extreme case of contact
interaction leading to the GP equation we do not have analytic solutions to argue our
case. However, the nonlinear term in this equation always vanishes on vortex lines. In
Section III we show that in the vicinity of the vortex line it remains much smaller than
the kinetic energy term. Therefore, we can argue that also in the case of contact
interactions the effect of the nonlinearity on the {\em qualitative} features of the
motion of vortex lines dynamics at close approach is negligible.

\section{Nonlinear Schr\"odinger equation with harmonic forces}

The starting point of our study is the nonlinear Schr\"odinger equation describing the
dynamics of atoms in a harmonic trap interacting via harmonic forces in the Hartree (mean
field) approximation. We shall assume that the trap is fully anisotropic with the
characteristic frequencies denoted by $\tilde\omega_x, \tilde\omega_y$ and
$\tilde\omega_z$. The interatomic harmonic forces can be either repulsive ($\Omega^2>0$)
or attractive ($\Omega^2<0$). In natural units ($\hbar = 1$, $m = 1$), our nonlinear
Schr\"odinger equation reads
\begin{eqnarray}\label{eqn1}
 i\partial_t\psi({\bf r}, t)
 &=& \left(-\frac{1}{2}\Delta + \frac{\tilde\omega_x^2x^2+\tilde\omega_y^2y^2+\tilde\omega_z^2z^2}
 {2}\right)\psi({\bf r}, t)\nonumber\\
 &-& \frac{\Omega^2}{2}\!\int\!
 d^3r'\vert\psi({\bf r}', t)\vert^2({\bf r}'-{\bf r})^2\psi({\bf r}, t).
\end{eqnarray}
Following our earlier work on systems with harmonic forces \cite{ibb,bb}, we introduce
the following set of time dependent global quantities ($N$ is time independent)
\begin{eqnarray}\label{global}
 N &=& \int\!\!d^3r\,\vert\psi({\bf r}, t)\vert^2,\\
 R_x(t) &=& \int\!\!d^3r\,x\vert\psi({\bf r}, t)\vert^2,\\
 P_x(t) &=& -i\int\!\!d^3r\,\psi^*({\bf r}, t)\nabla_x\psi({\bf r}, t),\\
 U_{xx}(t) &=& \frac{1}{2}\int\!\!d^3r\,x^2
 \vert\psi({\bf r}, t)\vert^2
 - \frac{R_x(t)R_x(t)}{2N},\\
 T_{xx}(t) &=&
 \frac{1}{2}\int\!\!d^3r\,\nabla_x\psi^*({\bf r}, t)\nabla_x\psi({\bf r}, t)
  -\!\frac{P_x(t)P_x(t)}{2N},\\
 W_{xx}(t) &=& -\frac{i}{4}\int\!\!d^3r\,\psi^*({\bf r}, t)
 (x\nabla_x + \nabla_x x)\psi({\bf r}, t)\nonumber\\
 && \;\;\;\;\;\; - \frac{R_x(t)P_x(t)}{2N},
\end{eqnarray}
and analogously for the components in the $y$ and $z$ directions. With the use of the
quantities ${\bf R}(t)$ and $U(t)$, Eq.~(\ref{eqn1}) can be rewritten in the form
\begin{eqnarray}\label{eqn2}
 i\partial_t\psi({\bf r}, t)
 &=& \left(-\frac{1}{2}\Delta
 + \frac{\omega_x^2x^2+\omega_y^2y^2+\omega_z^2z^2}{2}\right)\psi({\bf r}, t)\\
  &-& \Omega^2\left(U(t) - {\bf R}(t)\!\cdot\!{\bf r} + \frac{{\bf R}(t)\!\cdot\!{\bf
  R}(t)}{2N}
 \right)\psi({\bf r}, t),\nonumber
\end{eqnarray}
where $\omega_x^2 = \tilde\omega_x^2 - N\Omega^2,\;\omega_y^2 = \tilde\omega_y^2 -
N\Omega^2,\;\omega_z^2 = \tilde\omega_z^2 - N\Omega^2$ are the squared frequencies
modified by the mutual interaction and $U(t) = U_{xx}(t) + U_{yy}(t) + U_{zz}(t)$.

As a result of Eq.~(\ref{eqn1}), the global quantities obey a set of ordinary linear
differential equations (again we list only the $x$-components)
\begin{eqnarray}\label{ordeqn1}
 dN(t)/dt &=& 0,\\
 dR_x(t)/dt &=& P_x(t)\\
 dP_x(t)/dt &=& - \tilde\omega_x^2R_x(t),\\
 dU_{xx}(t)/dt &=& 2W_{xx}(t),\\
 dT_{xx}(t)/dt &=& -2\omega_x^2 W_{xx}(t),\\
 dW_{xx}(t)/dt &=& T_{xx}(t) - \omega_x^2 U_{xx}(t).
\end{eqnarray}
The solutions of these equations are
\begin{eqnarray}\label{soln}
 N(t) &=& N,\\
 R_x(t) &=& R_x(0)\cos(\tilde\omega_x t) +
  P_x(0)\frac{\sin(\tilde\omega_x t)}{\tilde\omega_x},\\
 P_x(t) &=& -R_x(0)\tilde\omega_x\sin(\tilde\omega_x t) + P_x(0)\cos(\tilde\omega_x t),\\
 U_{xx}(t) &=& U_{xx}(0)\cos^2(\omega_x t)\nonumber\\
  &+& T_{xx}(0)\frac{\sin^2(\omega_x t)}{\omega_x^2}
 + W_{xx}(0)\frac{\sin(2\omega_x t)}{\omega_x},\\
 T_{xx}(t) &=& U_{xx}(0)\omega_x^2\sin^2(\omega_x t)\nonumber\\
 &+& T_{xx}(0)\cos^2(\omega_x t) - W_{xx}(0)\omega_x\sin(2\omega_x t),\\
 W_{xx}(t) &=& -U_{xx}(0)\frac{\omega_x\sin(2\omega_x t)}{2}\nonumber\\
 &+& T_{xx}(0)\frac{\sin(2\omega_x t)}{2\omega_x} + W_{xx}(0)\cos(2\omega_x t).
\end{eqnarray}
The center of mass variables ${\bf R}(t)$ and ${\bf P}(t)$ oscillate with the original
trap frequencies $(\tilde\omega_x,\tilde\omega_y,\tilde\omega_z)$ and the variables $U,
T$, and $W$ that describe the motion of the internal quadrupole moment oscillate with the
doubled modified frequencies $(\omega_x,\omega_y,\omega_z)$.

We have found that each solution $\psi({\bf r}, t)$ of the nonlinear equation
(\ref{eqn1}) is related in a simple way to the corresponding solution $\phi({\bf r}, t)$
of the linear Schr\"odinger equation
\begin{eqnarray}\label{eqn3}
 i\partial_t\phi({\bf r}, t)
 = \left(\!-\frac{1}{2}\Delta + \frac{\omega_x^2x^2+\omega_y^2y^2+\omega_z^2z^2}{2}\right)
 \!\phi({\bf r}, t).
\end{eqnarray}
In order to prove this assertion we seek the solution of the nonlinear equation in the
form
\begin{eqnarray}\label{rel}
 \psi({\bf r}, t) = e^{i f(t) + i {\bf a}(t)\cdot{\bf r}}\;
 \phi({\bf r} - {\bf b}(t), t).
\end{eqnarray}
Upon substituting (\ref{rel}) into (\ref{eqn2}) we find that  $\phi({\bf r}, t)$ must
obey Eq.(\ref{eqn3}) while the scalar function $f(t)$ and the two vector functions ${\bf
a}(t)$ and ${\bf b}(t)$ must obey the following ordinary differential equations
\begin{eqnarray}
 d{\bf a}(t)/dt &=& \Omega^2{\bf R}(t) - {\bf b}(t),\label{ordeqn2}\\
 d{\bf b}(t)/dt &=& {\bf a}(t),\label{ordeqn3}\\
 df(t)/dt &=& \frac{1}{2}(\omega_x^2b_x(t)^2 + \omega_y^2b_y(t)^2 + \omega_z^2b_z(t)^2)\nonumber\\
 &-&{\bf a}(t)\!\cdot\!{\bf a}(t)
 - \Omega^2(U(t) + {\bf R}(t)\!\cdot\!{\bf R}(t)/N).\label{ordeqn4}
\end{eqnarray}
The solutions of these equations for $a_x(t)$ and $b_x(t)$ are
\begin{eqnarray}\label{soln1}
 a_x(t) &=& \frac{1}{N}\Bigl[-R_x(0)(\tilde\omega_x\sin(\tilde\omega_x t)
  - \omega_x\sin(\omega_x t))\nonumber\\
 &+& P_x(0)(\cos(\tilde\omega_x t) - \cos(\omega_x t))\Bigr],\\
 b_x(t) &=& \frac{1}{N}\Bigl[R_x(0)(\cos(\tilde\omega_xt) -
  \cos(\omega_x t))\nonumber\\
 &+& P_x(0)(\frac{\sin(\tilde\omega_xt)}{\tilde\omega_x}
  - \frac{\sin(\omega_x t)}{\omega_x})\Bigr],
\end{eqnarray}
and similarly for the remaining components. The phase $f(t)$ can be obtained from
(\ref{ordeqn4}) after performing a straightforward integration of products of
trigonometric functions. Next, we note that the vanishing of ${\bf a}$, ${\bf b}$, and
$f$ at $t = 0$ implies that the initial values $\psi({\bf r}, t = 0)$ and $\phi({\bf r},
t = 0)$ coincide. Therefore, we can argue that if $\phi({\bf r}, t)$ is a solution of the
initial value problem for the linear Schr\"odinger equation, then $\psi({\bf r}, t)$
given by Eq.~(\ref{rel}) will be the solution of the initial value problem for the
nonlinear Schr\"odinger equation. Note that the initial values of the wave function for
the nonlinear problem enter not only through the solution of the linear problem but also
through the initial values ${\bf R}(0), {\bf P}(0), U(0), T(0), W(0)$ of all global
variables.

The form (\ref{rel}) of the solution of the nonlinear problem shows that there are three
effects of the interatomic interactions. The wave function acquires a time- and
space-dependent phase factor, the frequencies of the trap are modified ($\tilde\omega \to
\omega)$, and the {\em whole wave function} undergoes an additional rigid motion
described by the vector ${\bf b}(t)$. This motion is made of oscillations with the
frequencies present in the problem. The amplitudes of these oscillations are determined
by the initial values of the center of mass position ${\bf R}(0)$ and momentum ${\bf
P}(0)$.

The relation (\ref{rel}) between the general solution of the nonlinear and the linear
Schr\"odinger equations obtained in our model with harmonic interatomic forces, enables
one to analyze the effect of this particular type of nonlinearity on the motion of vortex
lines. The phase factor $\exp(i f(t) + i {\bf a}(t)\cdot{\bf r})$ does not have any
influence on the vortex lines since it never vanishes. Thus, the only effect (apart from
the modification of the trap frequencies) is the shift of the argument in Eq.~(\ref{rel})
by ${\bf b}(t)$. This time-dependent shift causes oscillations of the same vortex
structure that is already present in the wave function $\phi({\bf r} , t)$ of the linear
problem. Thus, the topological structure of vortex lines is not affected by the
interatomic, harmonic interactions. Vortex creations, annihilations, reconnections, and
switchovers will occur unimpeded as in the linear case.

Eq.~(\ref{rel}) can also serve as a practical tool to generate explicit solutions of the
nonlinear Schr\"odinger equation (\ref{eqn1}) with (almost) arbitrary vortex structures.
To this end we only need to obtain such solutions of the linear equation. This can be
done with the help of the generating functions, as described in Ref.~\cite{bbs}.

As a simple illustration of this general analysis, let us consider the motion of a single
straight vortex line along the $z$-axis displaced at $t=0$ by $a$ from the origin in the
$x$-direction. We assume that the initial wave function has the form
\begin{eqnarray}\label{vort0}
 &&\phi_{\rm vort}({\bf r}, 0)\nonumber\\
 &=& N_0(x-a + iy)\exp(-\frac{\omega_x x^2+\omega_y y^2+\omega_z z^2}{2}),
\end{eqnarray}
where $N_0$ is the normalization constant and we have chosen the ground state as the
``background'' wave function. The time dependent solution of the linear Schr\"odinger
equation satisfying at $t = 0$ the initial condition (\ref{vort0}) is for all values of
$t$ given by
\begin{eqnarray}\label{vort1}
 &&\phi_{\rm vort}({\bf r}, t) =  N_0\exp(-i\frac{(\omega_x+\omega_y+\omega_z)}{2}t)\\
 &\times&\!(xe^{-i\omega_x t}\!\!-a + iye^{-i\omega_y t})
 \exp(-\frac{\omega_x x^2+\omega_y y^2+\omega_z z^2}{2}).\nonumber
\end{eqnarray}
One can obtain this solution using the general method introduced in Ref.~\cite{bbs}. The
position of the vortex line in the $xy$-plane is determined by the zeros of the equation
\begin{eqnarray}\label{vort2}
 \vert xe^{-i\omega_x t}\!\!-a + iye^{-i\omega_y t}\vert^2 = 0,
\end{eqnarray}
whose solutions are
\begin{eqnarray}\label{vort3}
 x = -a\cos(\omega_y t)/\cos((\omega_x-\omega_y)t),\\
 y = -a\sin(\omega_x t)/\cos((\omega_x-\omega_y)t).\nonumber
\end{eqnarray}
In the isotropic trap the zeros follow a circle, in all other cases they follow an
unbounded curve; the vortex line moves with an unlimited speed. In order to obtain the
solution of the nonlinear equation we need the displacement vector ${\bf b}(t)$. The only
nonvanishing components of ${\bf R}(0)$ and ${\bf P}(0)$ in this case are: $R_x(0) = a$
and $P_y(0) = a$ and the vector ${\bf b}(t)$ has only the $x$- and $y$-components
\begin{eqnarray}\label{b}
 b_x(t) &=& a(\cos(\tilde\omega_x t) - \cos(\omega_x t)),\\
 b_y(t) &=& a\left(\frac{\sin(\tilde\omega_y t)}{\tilde\omega_y} -
  \frac{\sin(\omega_y t)}{\omega_y}\right).\nonumber
\end{eqnarray}
The vector ${\bf b}(t)$ will draw a Lissajous figure in the $xy$-plane. The motion of the
vortex line in the nonlinear case in the anisotropic trap will be the composition of an
unbounded motion determined by the condition (\ref{vort2}) with the motion following the
Lissajous figure determined by Eq.~(\ref{b}).

\section{The role of the contact nonlinearity in the vicinity of vortex lines}

The dynamics of vortex lines in Bose-Einstein condensates in a symmetric trap is governed
by the GP equation
\begin{eqnarray}\label{eqn4}
 i\hbar\partial_t\psi({\bf r}, t)
 &=& \left(-\frac{\hbar^2}{2m}\Delta + \frac{m\omega {\bf r}^2}{2}\right)
 \psi({\bf r}, t)\nonumber\\
 &+& \frac{4\pi\hbar^2a}{m}
 \vert\psi({\bf r}, t)\vert^2\psi({\bf r}, t),
\end{eqnarray}
where $a$ is the scattering length. Although we cannot solve this equation analytically,
we can estimate the influence of the nonlinear term on the motion of vortex lines. Since
the wave function vanishes on the vortex lines, the influence of the nonlinearity is
weakening when one approaches the vortex line. On the other hand the interesting
topological properties of the vortex lines motion (reconnection, vortex creation, etc.)
occur at small distances. Therefore, we restrict our analysis to distances much smaller
(at least ten times) than the extension of the condensate. As an example we consider the
interaction of two perpendicular vortex lines separated by a distance $d$. The initial
wave function is assumed in the form
\begin{eqnarray}\label{two_lines}
 &&\psi({\bf r}, t=0)\\ &&= A(x+i(z-d/2))(y+i(z+d/2))\exp(-{\bf r}^2/L^2)\nonumber,
\end{eqnarray}
where $A = \sqrt{N}\pi^{-3/4}L^{-7/2}(3/2+d^4/16L^4)^{-1/2}$ is the normalization
constant, $L$ is the linear dimension of the condensate, and $N$ is the number of atoms
in the condensate. The ratio of the nonlinear part of the GP equation to the kinetic
energy part for this wave function is
\end{multicols}
\begin{eqnarray}\label{ratio}
 &&\frac{8\pi a\vert\psi\vert^2\psi}{-\Delta\psi}\\
 &&=\frac{32N\,a\,e^{-{\bf r}^2/L^2}(x^2+(z-d/2)^2)(y^2+(z+d/2)^2)(x+i(z-d/2))(y+i(z+d/2))}
 {\sqrt{\pi}L^3(3/2+d^4/16L^4)(8L^4+28L^2(x+iz)(y+iz)+d^2(3L^2-{\bf r}^2)
 -4{\bf r}^2(x+iz)(y+iz)+2id(x-y)(5L^2-{\bf r}^2))}.\nonumber
\end{eqnarray}
\begin{multicols}{2}
\noindent At the center this ratio is equal to
\begin{eqnarray}\label{ratio0}
\frac{8\pi a\vert\psi\vert^2\psi}{-\Delta\psi}\vert_{{\bf r}=0} = \frac{N\,a\,d^6}
 {\sqrt{\pi}L(L^2+3d^2/8)(24L^4+d^4)}.
\end{eqnarray}
\noindent Therefore it is small, of the order of $\xi=N(a/L)(d/L)^6$. Taking the linear
dimension of the trap $L = 5\times10^{-5}{\rm m}$, the vortex separation $d = L/10$, the
number of atoms $N=10^6$, and the scattering length $a=5\times10^{-9}{\rm m}$, one
obtains $\xi = 10^{-3}$. The same estimate of the ratio (\ref{ratio}) is valid for all
points lying at distances of the order of $d$ from the center. Next, we have to estimate
the role of the trap potential. The potential term modifies the shape of the wave
function for times comparable with the period of the trap $T\approx 10^{-2}{\rm s}$. The
characteristic time scale for the motion of vortex lines, as discussed in detail in
\cite{bbs}, is $T_0 = md^2/\hbar$. For the sodium atoms the value of this parameter is
$T_0\approx 3\times10^{-4}{\rm s}$. Therefore, the role of the trap potential is
negligible for small vortex separations and for times of the order of $T_0$. In
Ref.\cite{bbs} we used the linear Schr\"odinger equation to study the evolution of wave
function that was initially given by Eq.~(\ref{two_lines}). Vortex lines that initially
were separated by a distance $d$ crossed and reconnected after the time of the order of
$T_0$. In view of our present analysis the same behavior should be found also in the
evolution governed by the GP equation. The same arguments can be used in the case of more
complicated vortex structures provided we restrict ourselves to distances small as
compared to the linear dimension of the condensate and to times small as compared to the
trap period.

\section{Conclusions}

The aim of this study was to find out to what extent the topological properties of vortex
lines motion carry over from linear to nonlinear Schr\"odinger equations. We analyzed two
extreme cases of the nonlinear term that describes the mutual interaction of particles:
the harmonic interaction (extremely long range) and the contact interaction (zero range).
In the first case we were able to give a mathematical proof that the only change in the
motion of vortex lines is an overall, time dependent displacement of the whole vortex
line structure. In the second case we have shown that the features of the vortex lines
motion that occur at small distances and for short times are governed mainly by the
kinetic energy term in the Schr\"odinger equation. The influence of the nonlinear contact
term under those conditions is negligible. Thus, many features of the vortex lines motion
(especially those that involve a close approach of two vortex lines) are to a large
extent universal.

\end{multicols}

\end{document}